\documentclass[printer]{aa} 

\usepackage{graphicx}
\usepackage{lineno}
\usepackage{txfonts}
\usepackage{natbib}
\usepackage{color}
\usepackage{hyperref}

\usepackage[normalem]{ulem}

\begin{document}

\title{Diffuse Interstellar bands and ultraviolet extinction bump: a Milky Way perspective on distant galaxies}

\author{R. Lallement\inst{1}
}

\institute{LIRA, Observatoire de Paris, PSL University, CNRS, 5 Place Jules Janssen, 92190 Meudon, France\\
       \email{rosine.lallement@obspm.fr}
}

\date{Received ; accepted }
\titlerunning{DIBs and UV bump}
 
\abstract
{The  spectral width and the center wavelength of the ultraviolet (UV) absorption bump measured for at least two z$\simeq$7 galaxies were found to differ significantly from Milky Way (MW) values. A decrease of the width by $\sim$ 45\% and a positive shift of the center by $\sim$ 70-80 \AA~were measured. Within the MW,  the bump amplitude and width do vary; however, such a narrow bump has never been observed and no variability of the peak position has been convincingly found. On the other hand,  links have been recently detected between both the amplitude and the width of the bump and the strength of several Diffuse Interstellar absorption Bands (DIBs). The links were found to be limited to the so-called $\sigma$-type DIBs and their detection to be strongly favored if the data were limited to {\it monocloud}-type lines-of-sight (LOS), selected according to 3D maps of dust extinction.}
{We aim to extend the study of the links between MW DIBs and UV bump parameters to the center wavelength of the bump and to the ratio between the bump amplitude and the underlying continuum, and to compare the characteristics of the MW variations of the bump parameters to values at high redshift.}
{We used published catalogs of cross-matched  measurements of DIBs  and reddening law parameters. We assigned {\it monocloud} flags to all LOS. We focused on the strong $\sigma$-type DIBs at 5780 and 6284\AA~, and $\zeta$-type DIBs at 5797 and 5850\AA~ and searched for trends linking the bump parameters to the DIB strength normalized to the reddening.}
{Similarly to the case of the bump amplitude and width, the center wavelength of the bump is found to react to the abundance of $\sigma$-type DIB carriers and to be insensitive to the abundance of the $\zeta$-type DIB carriers, which dominate in dense and UV-shielded cloud cores. A strong abundance of $\sigma$ carriers induces a shift of the bump peak position to longer wavelengths and a decrease of its width. The variability range for these two parameters in the MW is about half the difference between average MW values and values in the distant galaxies. In the MW, an increase of  the abundance of $\sigma$ carriers also corresponds to an increase of the bump amplitude and of the ratio between the amplitude and the underlying continuum.}
{In the case of the MW, these results reinforce the hypothesis of the existence of individual types of hydrocarbon molecules simultaneously responsible for DIBs and part of the UV bump. They show that the majority of species responsible for narrow and positively shifted bumps in distant galaxies have a link with (or are) those producing the $\sigma$ DIBs and the long-wavelength part of the bump in the MW, and that, on the contrary, species producing the short-wavelength part of the bump in the MW are of a different nature and are absent along the paths to the regions of those distant galaxies that contribute most to the UV emission. These results are based on a limited number ($\simeq$ 95) of MW lines-of-sight. More measurements of the reddening curve obtained with a space borne UV spectrograph and more massive DIB measurements with ground-based spectrographs would bring better constraints on relationships between DIBs and UV bumps. Identifications of DIB carriers would shed light on their formation and the origin of the UV bump in MW and distant galaxies.}


\maketitle

\section{Introduction}\label{intro}

Accurate models of the spectral dependence of the interstellar extinction are mandatory tools to recover the intrinsic spectral properties of astrophysical sources of interest. Understanding the physical factors governing this spectral dependence is also particularly important with respect to dust emission properties since, e.g., a strong anti-correlation between the spectral slope of the extinction curve and the dust far infra-red emission spectral index $\beta$  has been observed \citep{Schlafly16,Zhang23}, with implications on the dust foreground to the CMB.  One of the main absorption features of the interstellar extinction curve is the so-called ultraviolet (UV) extinction bump centered around 2200 \AA~, whose origin is generally attributed to carbon nanoparticles. It is known to vary within the Milky Way in both strength and spectral width in significant proportions. In particular, the bump width is  found to depend on the environment. Dense media are associated with broad bumps, while more diffuse regions and regions of star formation have narrow bumps \citep[see, e.g. ][]{Gordon23}. On the other hand, a variability of the bump central wavelength is still a subject of debate \citep[see, e.g.][]{Blasberger17,Wang23}. Very recently, JWST/NIRSpec observations revealed positively wavelength shifted, very narrow and relatively strong UV bumps in the reddening curve of at least two galaxies at redshifts z=6.7 and z=7.1 corresponding to the epoch of re-ionization  \citep{Witstok23, Yang23_jades,Li24_jades,Lin25_jades, Ormerod25}. In particular, for the galaxy  JADES-GS-z6-0, \cite{Lin25_jades} measured 2263 \AA~and 250 \AA~ for the bump center and the FWHM, respectively.  The presence of a relatively strong bump in those galaxies devoid of the low and intermediate-mass AGB stars,  supposedly the main contributors to the interstellar dust production,  is challenging dust production models in the early universe \citep{Witstok23, Ormerod25}.  Independently of the strength, the origin of the shifted bump center and of its narrowness is under debate. \cite{Li24_jades} found that small graphitic grains can not be responsible for such a bump shape and position, while \cite{Lin25_jades} showed that the combination of small and large PaHs could produce a bump location and shape similar to those observed. 

Diffuse Interstellar Bands (DIBs) are numerous \citep[$\sim$600 detected, see, e.g., ][ and references therein]{Fan19}, weak, irregular absorptions imprinted in the spectra of all astrophysical sources located behind interstellar clouds. DIBs are generally attributed to large carbonaceous molecules or molecular ions in the gaseous phase due to their wavelength range, multiple-peak sub-structures and absence of polarization  \citep{Sarre95, Ehrenfreund96, Salama96, Cox2011b, CamiCox14_IAU, Kwok23}. Only one DIB carrier, the buckminster fullerene cation C$_{60}^{+}$  has been identified \citep{Campbell15,Cordiner19}.  DIBs have been detected in the Magellanic clouds and distant galaxies \citep{Heckman2000, Ehrenfreund2002, Cordiner2006, Welty2006, Cordiner11, Monreal15}.

DIBs whose EW ratios do not vary strongly from one LOS to another have been classified in {\it families}. Two of these {\it families} bring together DIBs which show the same response to the intensity of the UV radiation impacting the contributing cloud: the so-called $\sigma$ DIBs tend to disappear in dense regions shielded from UV radiation, while $\zeta$ DIBs remain unaffected or are slightly affected by such a shielding \citep[see][ and references therein for more details]{Vos11, Fan17, Fan22}. The names $\sigma$ and $\zeta$ come from their prototype sightlines towards the star $\sigma$Sco and $\zeta$Oph \citep{Krelowski87}. Apart from the links to the quantity of intervening matter and to the UV radiation field, other factors are impacting on the ratio between the DIB strength and the reddening, as shown by \cite{Ensor17}, however, their nature is still unknown.

Links between DIB carriers and the UV bump have been searched in different ways. Null \citep{Benvenuti89}, marginally significant \citep{Witt1983} or positive \citep{Desert1995, Megier2005} correlations between DIB equivalent widths (EWs) normalized to the reddening E(B-V) and normalized bump strengths E(bump)/E(B-V) were obtained based on a small number of target stars and for some strong optical DIBs, followed by inconclusive results based on more targets and more DIBs  \citep{Xiang11, Xiang17}. More recently, 3D maps of Milky Way dust extinction built thanks to Gaia and 2MASS have been used to revisit the link \citep[][hereafter LVC]{Lall24}. The idea was that if clouds with different properties and ratios between DIB strengths and UV bump parameters are distributed along a line of sight (LOS), the resulting ratio measured for the whole LOS is an average of their properties, and this multiplicity prevents the detection of any variability. To avoid such a smearing, it is necessary to distinguish between LOSs dominated by a unique cloud and LOSs to which several clouds contribute. To do so, the 3D maps were used to separate the measurements into two groups, those characterized by the dominance of one single dense cloud or a group of clouds close to each other (FLAG 1 LOS) and those with extinction distributed in structures far away from each other (FLAG 0 LOS). In the case of FLAG 1, a positive (respectively negative) link between the UV bump height (respectively width) of the extinction curve and the abundance of the so-called {\it sigma} ($\sigma$)  DIBs has been found, while {\it zeta} ($\zeta$) DIBs failed to reveal such correlations (LVC). A proxy for the UV bump height based on the strength of a few strong DIBs was proposed by the authors. 

In the above-mentioned work, no attention was paid to the bump central wavelength. The recent discoveries of a wavelength shift between the peak position for the two distant galaxies and the one of the MW motivated the present study. We used the extensive catalog of \cite{Xiang17} and their very useful compilation of reddening curve parameters and DIB measurements.  We focused on four strong DIBs, two of $\sigma$ type, and two of $\zeta$ type. We considered the same flags as in LVC based on 3D-maps and searched for links between DIB strengths and UV bump peak location. Section 2 describes the dependence of the bump center location, bump width, bump amplitude and ratio between the bump amplitude and the underlying continuum on the  ($\sigma$)  DIB abundance, suggestive of a similarity between DIB and bump carriers.  We confirm the absence of biases induced by normalizations by means of a study of the link between the bump parameters and the unit-less, often-used ratio between the equivalent widths (EWs) of the prototype $\sigma$ and  $\zeta$ DIBs at 5780 and 5797 \AA~, EW(DIB 5780) /EW(DIB 5797).  In section 3 we extrapolate the ranges of peak wavelength, width, amplitude, and amplitude to continuum ratio to large 5780\AA~  DIB abundances, compare with the bump parameters in z=$\sim$7 galaxies and conclude.

\section{UV bump parameters and DIBs}\label{indivdib}

In the MW, the optical thickness of even the strongest DIBs  remains less than unity, therefore the equivalent width (EW) of a given absorption band can be used as a measurement of the column of the corresponding carrier distributed along the LOS.  Because EWs of DIBs are correlated with the amount of interstellar matter distributed along the LOS, and, as a result, with the extinction (or the reddening), as shown by the remarkable similarity between 3D maps of dust extinction and 3D maps of the so-called Gaia DIB at 8620\AA~ \citep{Cox24}, quantities able to trace a variability of the physical properties of the intervening interstellar medium are extinction- or reddening- normalized EWs, i.e. EW/E(B-V) or EW/A$_{V}$ \citep{Witt1983}.  \cite{Xiang11}, then \cite{Xiang17} compiled DIB EW measurements from the literature for all stars with existing measurements of the LOS extinction curve. When several measurements do exist for the DIBs, the authors estimated either the most accurate value or a weighted average.  \cite{Xiang17} provided in their Table 1 the coefficients of the decomposition of the extinction E($\lambda$-V)/E(B-V) as a function of 1/$\lambda$ (in $\mu^{-1}$)  into three components, a linear background c1 + c2 * x, a Drude profile for the UV bump c3 * D ( x : $\gamma$, x$_{0}$) and a far-UV (FUV) rise expressed as a polynomial, following the classical scheme of \cite{FitzpatrickMassa90}. x$_{0}$ is the peak position of the bump, and $\gamma$ is its FWHM. Here we are interested in the first two terms only. In this formalism the UV bump amplitude is c3/$\gamma^{2}$, and the underlying continuum is c1 + c2 * x$_{0}$. For an extinction law expressed as A($\lambda$)/A$_{V}$, the coefficients x$_{0}$ and $\gamma$  are similar, the bump amplitude is c'3 /$\gamma^{2}$ with c'3= c3/R$_{V}$ and the underlying continuum is c'1 + c'2 * x$_{0}$ with c'1=c1/R$_{V}$ +1 and c'2=c2/R$_{V}$. For the bump amplitude and the ratio between the bump and the underlying continuum, we will use this second formalism A($\lambda$)/A$_{V}$. The  flags associated to the types of LOS, {\it monocloud}-type or not, are listed in LVC.

\begin{figure}
  \centering
 \includegraphics[width=0.49\textwidth]{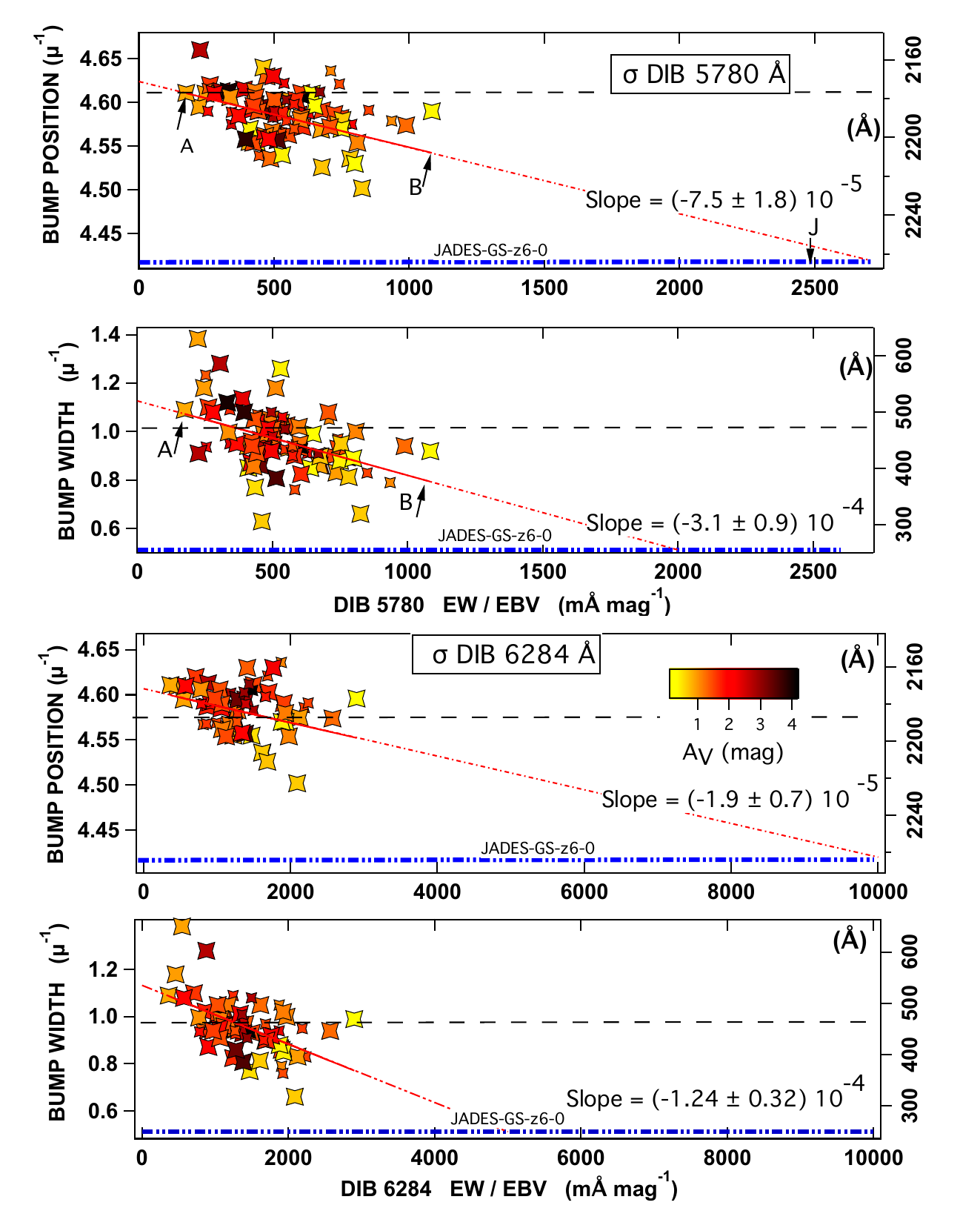}
   \caption{Position and width (FWHM)  of the UV bump of the extinction law A($\lambda$-V)/E(B-V) expressed as a function of 1/$\lambda$ in $\mu^{-1}$, here plotted as a function of the EW normalized to the reddening E(B-V)  of the $\sigma$-type DIBs at 5780 \AA ~ (Top graphs) and at 6284 \AA ~ (Bottom graphs) . The size of the marker is smaller for lines-of-sight flagged as "multiclouds" based on 3D dust maps (Flag 0, see text). The linear fits are restricted to "monocloud"-type lines-of-sight (Flag 1). They are extrapolated to reach the values measured for the galaxy JADES-GS-z6-0, indicated by thick horizontal dash-dot lines. An approximated corresponding scale in \AA~units is given on the right. It is interpolated linearly between the mean Milky Way values and the JADES measured values. Average Milky Way values are indicated by horizontal dashed lines. Both bump position and width are found to decrease when the abundance of the DIB carrier is increasing. The color scale refers to the dust extinction A(V) along the line-of-sight from the Sun to the target star. It allows to identify potential biases related to the amplitude of the LOS extinction.}
 \label{X0_GAMM_5780_6284EBV}
\end{figure}

\begin{figure}
  \centering
 \includegraphics[width=0.495\textwidth, height=11cm]{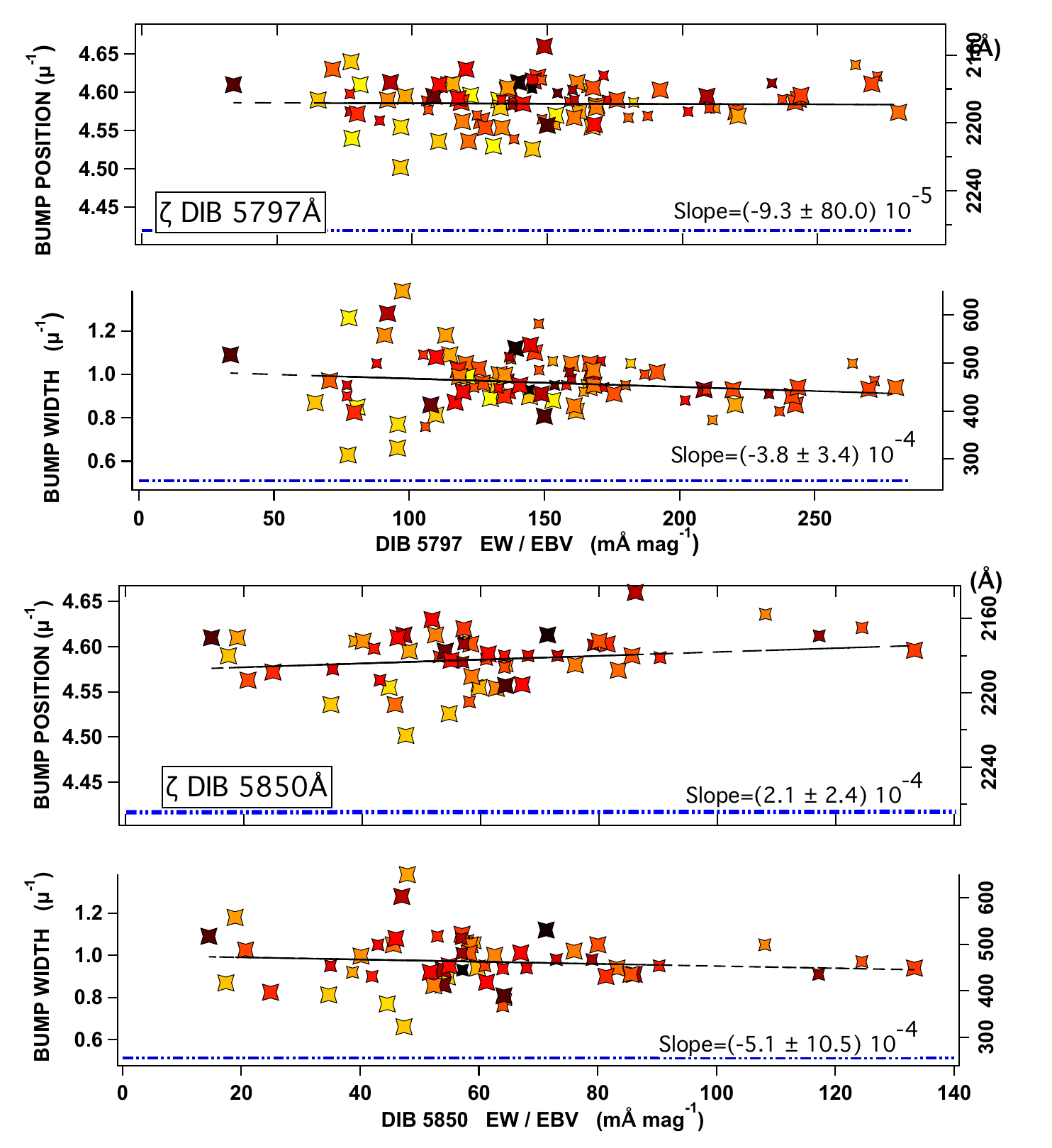}
  \caption{Same as Fig. \ref{X0_GAMM_5780_6284EBV} for the $\zeta$-type DIB at 5797 \AA~ and 5850\AA~ . At variance with the two $\sigma$-type DIB at 5780 and 6284 \AA ~, there is no detected trend.}
 \label{X0_GAMM_5797_5850EBV}
\end{figure}

We show in Fig.  \ref{X0_GAMM_5780_6284EBV} the UV bump position x$_{0}$  and the bump width $\gamma$ as a function of the  DIB EW  normalized to the reddening EW/E(B-V) for the two strong DIBs at 5780\AA~ (94 LOS) and  6284\AA~ (64 LOS).  Those two DIBs are well known as members of the $\sigma$ family \citep[see, e.g., ][]{Vos11, Ensor17}. The 6284\AA~ DIB has less data points, since its measurement requires a careful correction for the telluric O$_{2}$ absorption \citep{Puspitarini13}. All measurements are shown; however, we retained for the fit monocloud-type LOS only (59 LOS from a total of 95, marked by larger signs in the Fig. \ref{X0_GAMM_5780_6284EBV} ). One of the four graphs (bump width for DIB 5780) was already presented in LVC; however, we included it here again for completeness and for the extrapolation discussed below.  The data were fitted to a linear relationship using a classical Levenberg-Marquardt algorithm.  Here,  we did not use any errors on normalized equivalent widths EW/E(B-V) nor on dust curve parameters. Uncertainties on the EWs are available, but uncertainties on E(B-V) and on the dust curve parameters were not. Instead,  the algorithm estimated error intervals on the slope and intercept of the fitted linear curve by assuming that there is no error on the normalized EW and that the uncertainty on x$_{0}$ (resp. $\gamma$) is the same for all data points and equal to the average dispersion around the linear fit. It is important to note that the correlations can not be tight, essentially due to the absence of perfect FLAG 1 cases. As a matter of fact, LOS with a unique cloud containing the quasi-totality of the absorbers are rare. Even in the case of a single cloud along the LOS, its external and central regions differ in ambient radiation field, ionization state and and density, and this is expected to smear out in some extent DIB differences associated with these parameters. As a result, only trends are expected.  Here, our criterion for the existence of a relationship is based on the ratio between the uncertainty on the slope and the slope itself. Both values are indicated on each graph. Based on this criterion, it can be seen that for these two $\sigma$ DIBs, there is negative dependence of the bump center location and bump width (both expressed in inverse wavelength $\mu^{-1}$ units) on the DIB strength. The relative uncertainty on the slope is between 24 and 37\%, showing a relationship at 3 to 4 sigma confidence level. Fig. \ref{X0_GAMM_5797_5850EBV} shows the same parameters x$_{0}$ and $\gamma$, this time for the  5797  and 5850\AA~  DIBs. Those two are known for being unattenuated in UV-shielded, dense clouds, i.e. as members of the $\zeta$ family. It can be seen that there are no detected trends for such DIBs, with uncertainties on the slopes between equal, above or far above the slopes themselves. The decrease of the bump width in  LOS  with large $\sigma$ DIB abundance, i.e. crossing diffuse areas and avoiding opaque molecular clouds, is in good agreement with the bump width trend found by \cite{Gordon23}, based on other markers of the ISM type than the DIBs. In the case of the two $\sigma$ DIBs (Fig.  \ref{X0_GAMM_5780_6284EBV}), we have extended the figure and the linear fit far enough to reach the bump location and bump FWHM measured by \cite{Lin25_jades} for the galaxy JADES-GS-z6-0, namely 2263 \AA~and 250 \AA~respectively, (a correspondence between wavelengths in \AA~ and inverse wavelengths in $\mu^{-1}$ units is shown in the figures). Those JADES-GS-z6-0 values are indicated by horizontal lines.  These values are reached for normalized EWs of the DIBs between 2 to 4 times larger than the largest value of the dataset (about 2000 to 2800 m\AA~mag$^{-1}$ for DIB 5780 and 4000 to 10000 m\AA~mag$^{-1}$ for DIB6284.

%

\begin{figure}
  \centering
  \includegraphics[width=0.49\textwidth]{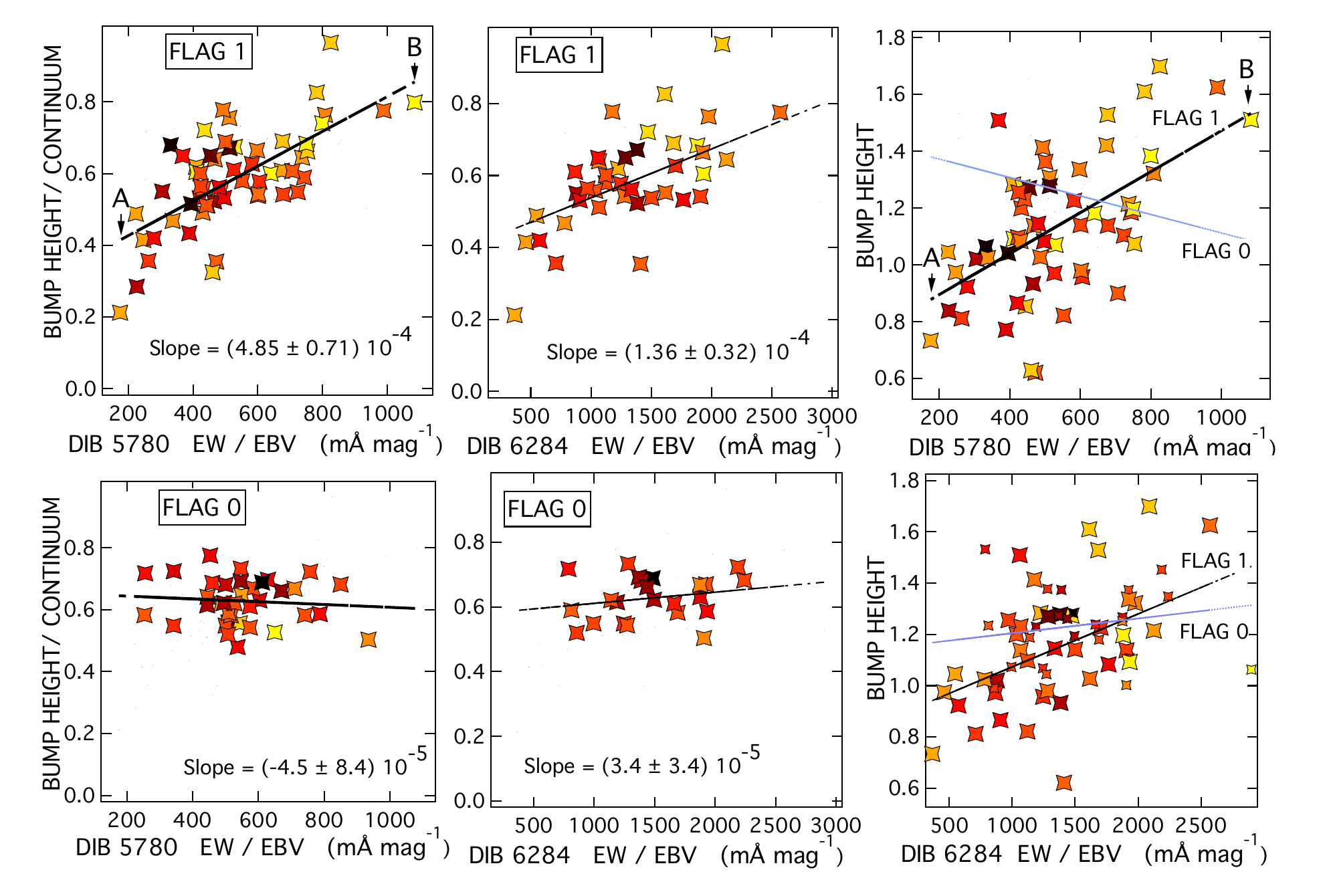}
  \caption{Amplitude of the bump. Left and Middle: ratio between the bump amplitude and the underlying continuum as a function of the 5780 and 6284 \AA~ DIB equivalent width normalized to the reddening. The values correspond to the A($\lambda$)/A$_{V}$ extinction curve. The color scale refers to the LOS extinction A$_{V}$ as in Figures \ref{X0_GAMM_5780_6284EBV} to \ref{X0_GAMM_5797_5850EBV}. Note the difference between FLAG 1 ((top) and FLAG 0 (bottom)  lines-of-sight. Right: Amplitude of the bump for the two DIBs: the fitted linear relationships are for Flag 1 LOS (black line) and Flag 0 LOS (pale blue line). The corresponding slopes are  7.2 ± 1.4 10$^{-4}$,  -3.2 ± 1.8 10$^{-4}$, 2.1 ± 0.6 10$^{-4}$,   5.7 ± 7.2 10$^{-5}$ for 5780/FLAG1, 5780/FLAG0, 6284/FLAG1, 6284/FLAG0.}
 \label{BUMPSURSOCLE_5780}
\end{figure}

The above results show that, in addition to the bump amplitude and bump width, the position of the bump peak is also related to the abundance of the $\sigma$ DIBs. We pursued our study by considering the relative amplitude of the bump  with respect to the underlying continuum (A$_{r}$(bump peak) - A$_{r}$(continuum))/ A$_{r}$(continuum). As said above, we use for that the A$_{r}$= A($\lambda$)/A$_{V}$ formalism for the extinction law.   Fig. \ref{BUMPSURSOCLE_5780}  represents the bump relative amplitude as a function of the reddening-normalized DIB EWs for the same two $\sigma$ DIBs. Here we used separated  graphs for the  FLAG 1 and FLAG 0 LOS. The values of the slopes and uncertainties of the linear fits are indicated on each graph. They show that, again, there is a detected positive relationship for FLAG 1 for both $\sigma$ DIBs, an absence of trend for the DIB 5780 and FLAG 0, a small  negative trend (relative uncertainty on the slope of 56 \%) for the DIB 6284 and FLAG 0. The dispersion around the linear fits is larger than in the case of the bump  position and width, a likely result of the stronger difficulties in measuring amplitudes. We also show in Fig. \ref{BUMPSURSOCLE_5780} the height of the bumps (A$_{r}$(bump peak) - A$_{r}$(continuum)) as a function of the reddening-normalized  DIB EWs. For FLAG 1 lines-of-sight, there is an increase of the bump amplitude. As a conclusion, in the MW,  increasing the proportions of $\sigma$ DIBs simultaneously shifts, narrows, and amplifies the UV bump.

\begin{figure}
  \centering
 \includegraphics[width=0.495\textwidth, height=6cm]{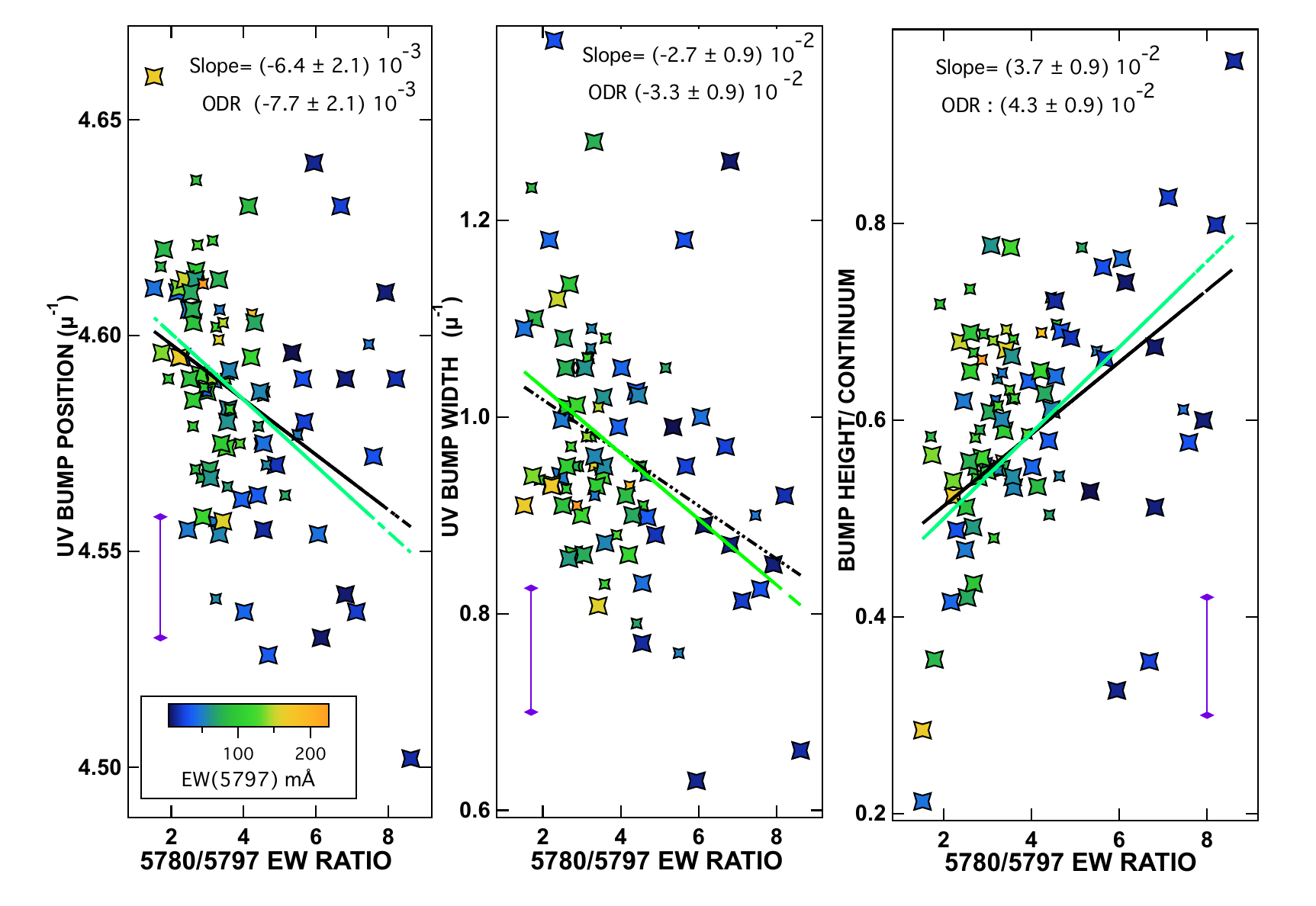}
   \caption{Position, width and height-to-continuum ratio of the UV bump, here as a function of the unit-less ratio between the 5780 \AA~ DIB equivalent width and the one of the 5797 \AA~ DIB. The color scale refers to the EW of the 5797 \AA~. It can be seen that the scatter around a mean trend becomes higher for low 5797 \AA ~ DIB EWs, and reflects the growing uncertainty on the  ratio. The black lines are for the linear fit applied to all stars with FLAG 1, as described in the text. The green lines are the ODR fits results (see text).}
 \label{X0_GAMM_HEIGHT_UVRATIO}
\end{figure}

One could argue that we have been  comparing quantities that are all normalized to the reddening (or the extinction), i.e. both the parameters of the extinction law or normalized DIB EWs.  It is known that, in such a case, some correlations may appear that are simply due to errors on the common value of the denominator in the used ratios.  Although it is hard to imagine how it could introduce a link with the bump position, or its width, we also considered the unit-less and reddening-independent  quantity EW(5780)/EW(5797), the ratio between the equivalent widths of a strong $\sigma$ DIB and the one of a strong $\zeta$ DIB, as an additional check of the DIB-bump link. Fig. \ref{X0_GAMM_HEIGHT_UVRATIO} shows the dependence of the bump position, width, and ratio between bump amplitude and underlying continuum as a function of this ratio, and a linear fit similar to previous ones. One difficulty here is the weakness of the 5797 \AA~DIB, and the resulting large uncertainty on the EW(5780)/EW(5797) ratio in the case of the smallest values of EW(5797). In order to take into account the large dispersion of the data points at these low values, we additionally performed an orthogonal distance regression (ODR) fit using errors on the two quantities that are compared. For the uncertainty on the EW(5780)/EW(5797)  ratio, we used the individual measurements and errors of the two DIBs  and added relative errors quadratically. For the error on the bump parameters, we assumed a unique value on the order of the dispersion around the linear fit and used an iterative method. I.e., we started with a {\it first guess} initial value, performed the fit, computed the dispersion, adopted it as the uncertainty, fitted again, etc.., until the input value and the dispersion around the ODR linear fit became identical. This uncertainty is indicated in the figures by violet vertical lines. It can be seen that the ODR fits produce a slightly larger slope than the other method. In any case, the main result is the confirmation of the three trends, i.e., increasing the dominance of the $\sigma$-type DIB 5780\AA~ with respect to the $\zeta$-type DIB 597\AA~  has the result of shifting the bump center to lower 1/$\lambda$ (i.e., shifting the bump center to longer wavelengths), decreasing the bump width and increasing its amplitude w.r.t. the continuum.

\section{From MW to distant galaxies}\label{evolbump}

There are two aspects in the above results. The first one is the reinforced relationship between the abundance of the carriers of some strong $\sigma$ DIBs, i.e. free-flying individual species in the gaseous phase, and the shape of the extinction bump in MW lines-of-sight. This is suggestive of a strong similarity between $\sigma$ DIB and bump carriers, and maybe of their identity. A shift of the bump center on the order of 30 \AA~and a decrease of the width by $\sim$35\% (see Fig \ref{X0_GAMM_5780_6284EBV}) suggest important changes in the relative abundances of the bump carriers. We note that this is in line with the recent model results of \cite{Lin23} who showed that, if the bump carriers are peri-condensed PaH molecules and/or ions, then the bump width is not compatible with a mixture of molecules, cations, or anions in a large range of sizes, and that instead specific sizes and structures are required.  The effect of the selection of mono-cloud lines-of-sight on the appearance of a DIB-bump link can be understood if one considers that there are different  phases in the intervening ISM, producing different bump shapes, but that complex and/or large distributions of clouds with these different characteristics will result in a smearing and the disappearance of any variability. Indeed, average values of the bump parameters for the multi-cloud lines of sight fall in the middle of the ranges found for the mono-cloud type. The reason for the disappearance of any link with the bump if one used $\zeta$ DIBs is an interesting question. $\zeta$ DIB carriers are only very likely attenuated in dense, UV-shielded regions, where $\sigma$ DIBs tend to disappear. Potential explanations are that those regions do not contribute significantly to the bump. 

\begin{figure}
  \centering
 \includegraphics[width=0.49\textwidth, height=8cm]{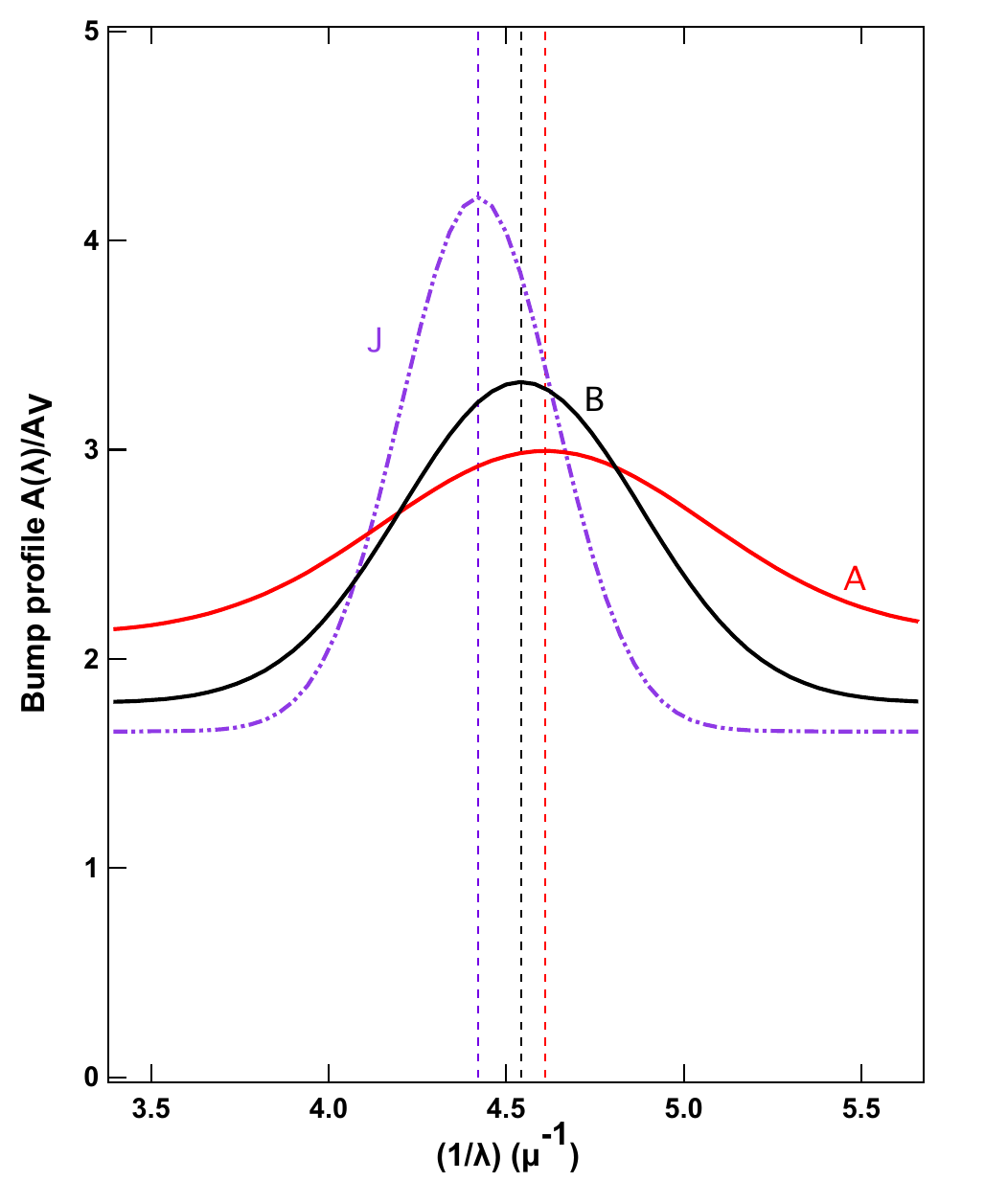}
   \caption{Simplified Gaussian profiles of the bump for different UV bump parameters. The positions and widths of the red (resp. black) profiles correspond to the extrema of the fitted linear relationships in Fig. \ref{X0_GAMM_5780_6284EBV}, marked by letters A (resp. B) in the two top graphs, corresponding to the 5780 \AA~DIB. The bump height and the continuum were derived from fitted bump height and bump-over-continuum relationships for the same DIB (see text).  The position and width of the violet profile  are those of the JADES-GS-z6-0 galaxy. The bump height and the continuum were derived by extrapolation of the fitted linear relationships to the value EW(5780)/EBV = 2500 (point J in top graph of Fig \ref{X0_GAMM_5780_6284EBV}). }
 \label{GAUSS}
\end{figure}

The second, and striking aspect of the above trends is that both the position and the width of the bump are evolving in the MW in a way to approach those measured in the above-mentioned high-z galaxies along lines-of-sight for which the $\sigma$ DIBs are getting increasingly strong relatively to the extinction and dominate in an increasing way the $\zeta$ DIBs. We illustrate this continuity in Fig. \ref{GAUSS}. We used the bump locations, bump widths, bump amplitudes (in A($\lambda$/A$_{V}$ units) and bump/continuum ratios corresponding to the extremities of the linear fits of Fig. \ref{X0_GAMM_5780_6284EBV} (top two graphs) and \ref{BUMPSURSOCLE_5780} (top left and right graphs). These minimum and maximum average values of MW fitted parameters are indicated in the four corresponding graphs by the letters A and B. For each set of A and B values, we constructed the corresponding bumps A and B of Fig. \ref{GAUSS}, assuming the bump has a Gaussian profile and is above a  flat continuum. Superimposed is a third Gaussian profile (letter J) corresponding to the position and width measured for JADES-GS-z6-0 galaxy. For this simplified model, we used arbitrarily as bump height and height to continuum ratio the values of the linear fits extrapolated to the high value of 2500 m\AA~mag$^{-1}$ for the 5780\AA~DIB normalized EW (Points J on Fig. \ref{X0_GAMM_5780_6284EBV}, top graph). From the point of view of position and width, it can be seen that the extrapolation of the trends observed in the MW (A to B) may lead to a profile as observed for this galaxy. On the other hand, the extrapolated bump height, with A(bump)/A$_{V}$ on the order of 2 is not compatible with the actual measured amplitude A(bump)$\sim$0.43 mag and E(B-V) $\sim$ 0.25 for JADES-GS-z6-0. The same is true for the relative amplitude with respect to the continuum, if R$_{V}$ is in a range of classical values. Here, however, it is important to distinguish between what is measured in the case of the distant galaxies, i.e. an attenuation curve affecting the integrated light of contributing sources, each of them being absorbed by a different interstellar medium, and MW measurements of actual extinction/reddening curves of individual sources, corresponding to a unique pathway for a given target. For galaxies, the way clouds and contributing UV-bright stars are distributed in 3D space and the geometry of the observation have a strong effect on the attenuation curve, as it has been modeled in detail \citep{Witt2000}. Moreover, the dust abundance and properties in the MW and in this re-ionization epoch galaxy are certainly extremely different and comparisons of their bump amplitudes may not be relevant.  As detailed by \cite{Ormerod25} the relative amplitude of the bump may be highly variable in cosmic dawn galaxies and depends on the spatial distributions of the star-forming regions and of regions affected by recent mergers, if present. \cite{Markov25} have analyzed the attenuation curves of a large sample of galaxies with z between 2 and 8 observed with JWST, and found that the UV bump amplitude is globally decreasing with z, but may be above the one of the MW in several high-z objects. In the future, it would be interesting to determine whether those strong bumps are narrow and shifted in the same way as in the case of JADES-GS-z6-0.  Still, the trends detected here suggest that species contributing to the short-wavelength part of the MW bump become under-abundant in regions characterized by strong $\sigma$ DIBs and are absent from the JADES-GS-z6-0 galaxy, and, reciprocally, that species contributing to the long-wavelength part of the bump in this distant galaxy have decreasing abundance in MW regions where $\sigma$ DIBs are under-abundant. It is tempting to speculate that $\sigma$ DIB carriers may be direct products of dying stars which have not been processed in collapsing clouds and are already abundant in high redshift galaxies, while $\zeta$ DIB carriers are secondary species produced in star-forming regions and are less abundant in those galaxies. Following this scheme, in the MW, the former species dominate in diffuse, low density regions while the latter dominate in opaque clouds close to (or within) star-forming regions. However, the identification of both types of DIB carriers thanks to laboratory dedicated experiments, theoretical calculations, and detailed photochemical models are required to draw conclusions and to progress in our understanding of the evolutionary processes at play. This study is based on a small number of target stars, due to the limited number of DIB measurements on one side, and UV extinction curves on the other hand. A UV spectrograph in Earth's orbit would bring massive amounts of extinction curves of MW early-type stars, for which present and future ground-based high resolution spectrographs would provide the complementary DIB measurements. Such UV and optical  spectra would be of strong interest in the context of the large collection of revolutionary data on distant galaxies gathered by the JWST.


\begin{acknowledgements}

I thank the referee, A. Witt, for his very helpful comments and suggestions. This research has made use of the VizieR catalog access tool, CDS, Strasbourg, France (DOI: 10.26093/cds/vizier). The original description of the VizieR service is published in 2000, A\&AS 143, 23.

\end{acknowledgements}

\bibliographystyle{aa}
\bibliography{dibuvbumprev2}

\end{document}